\documentclass[prl, twocolumn, amsmath, showpacs]{revtex4}

\usepackage{graphicx}
\usepackage{amssymb}

\begin{document}
\title{Absence of pre-classical solutions in Bianchi I loop quantum cosmology}
\author{Daniel Cartin}
\email{cartin@naps.edu}
\affiliation{Naval Academy Preparatory School, 197 Elliot Street, Newport, Rhode Island 02841}

\author{Gaurav Khanna}
\email{gkhanna@UMassD.Edu}
\affiliation{Physics Department, University of Massachusetts at Dartmouth, North Dartmouth, Massachusetts 02747}

\date{\today}
\begin{abstract} 
Loop quantum cosmology, the symmetry reduction of quantum geometry for the study of various cosmological situations, leads to a difference equation for its quantum evolution equation. To ensure that solutions of this equation act in the expected classical manner far from singularities, additional restrictions are imposed on the solution. In this paper, we consider the Bianchi I model, both the vacuum case and the addition of a cosmological constant, and show using generating function techniques that only the zero solution satisfies these constraints. This implies either that there are technical difficulties with the current method of quantizing the evolution equation, or else loop quantum gravity imposes strong restrictions on the physically allowed solutions.

\end{abstract}
\pacs{04.60.Pp, 04.60.Kz, 98.80.Qc}

\maketitle

When a new method of looking at physical situations arises, it is frequently too complicated to look at the general case, and thus it is necessary to look at simpler versions as test models. For example, in quantum mechanics one considers the hydrogen atom before examining more involved systems. These reduced models not only are easier to solve, but frequently provide information about what to expect from the more generic situation. This is the case with loop quantum cosmology~\cite{lqc}, a symmetry reduced version of the full theory of quantum geometry~\cite{rov} (also known as loop quantum gravity), which is intended to provide a framework joining together general relativity and quantum mechanics.

Much like the Bohr quantization condition for the electrons of the atom, space-time itself is no longer continuous in quantum geometry, but instead is discrete. Since this means the states representing geometry are also discrete, the quantum evolution equation becomes a difference equation. In general, however, solutions to a recursion relation will oscillate in sign as one moves in a particular direction in the parameter space. Ordinarily, unphysical states like these would be eliminated by the imposition of an inner product, but this is unavailable at the moment in loop quantum cosmology. To get around this, a notion of {\it pre-classicality}~\cite{boj01} is used to pick out those quantum solutions to match a semi-classical state far away from the singularity. This condition will eliminate all states which vary greatly at Planck-length scales. We will see here how the pre-classicality of a solution can be imposed as a global property.

Not only does loop quantum cosmology allow us to study simple situations of physical interest -- for example, the quantized version of the isotropic Friedman-Robertson-Walker (FRW) space-times that serves as the basis of modern cosmology -- so that physical predictions can be made, but it can give information about the full theory. In particular, it is closer in spirit to full quantum geometry than the minisuperspace models considered in standard quantum cosmology would be to a quantized general relativity. For the latter, the metric was first reduced to a specific model, then the equations of motion were quantized. Loop quantum cosmology takes the opposite path: starting with the kinematic space of quantum states from the full model, those states that obey the particular symmetry are chosen and then the Hamiltonian constraint on these states is quantized.

So far, most work on loop quantum cosmology has been done on isotropic models, looking at both quantum and semi-classical aspects. This has included a resolution of the classical singularity~\cite{boj02}, cosmological applications such as inflation~\cite{inf} and oscillatory universes~\cite{osc}, and semi-classical approximations of the quantum Hamiltonian constraint~\cite{semi}.  However, some anisotropic models have also been studied, including Bianchi IX~\cite{boj-dat-hos04} and the Bianchi I model~\cite{car-kha-boj04} when a local rotational symmetry (LRS) is assumed. This allows us to see how loop quantum cosmology fares in more general settings, while still being simple enough to get exact solutions to the quantum Hamiltonian constraint. In particular, for Bianchi I LRS, generating function techniques~\cite{wil93} were used to find possible solutions to this constraint, a partial difference equation in two parameters. By placing boundary conditions on the differential equation associated with this recursion relation, one could find all pre-classical solutions to the original quantum evolution constraint. In this paper, we examine the full Bianchi I model and carry out the same program (this model has also been looked at using related methods~\cite{mal04}). Yet, as we will see, the result we reach will be a negative one -- the only pre-classical solutions are the null solutions.

The Bianchi I  model is one of the simplest anisotropic models, with a metric of the form $ds^2 = - dt^2 + \sum_i a_i^2 (t) dx_i ^2$ $(i=1,2,3)$, where the functions $a_k (t)$ are the scale factors in the respective spatial direction. The flat $(k=0)$ FRW metric used in cosmology is the special case where the three scale functions are identical. In the connection formalism~\cite{boj03}, there are nine invariant degrees of freedom, which are reduced to six after the gauge freedom is removed. When restricted to diagonal metrics, we have that the connection 1-form is given by $A^i _a = c_{(K)} \Lambda^i _K \omega^K _a$, where (because of the diagonalization) $\omega^K _a$ are a {\it fixed} set of left-invariant 1-forms, and $\Lambda^i _K \in SO(3)$ is a rotation matrix. The momenta conjugate to the coefficients $c_{(K)}$ are given by the components of an invariant densitized triad $E^a _i = p^{(K)} \Lambda^K _i X^a _K$, where the $X^a _K$ are the vector fields dual to $\omega^K _a$. These triad  components are related to the scale factors $a_i$ of a Bianchi I metric by $p_1 = | a_2 a_3 | sgn(a_1)$, and so on for the other two momenta. The symplectic structure given by $\{c_I, p^J\} = \kappa \gamma \delta^J _I$, where $\gamma$ is the Barbero-Immirzi parameter of loop quantum gravity and $\kappa = 8 \pi G$.

We now briefly review the quantization of the Bianchi I model~\cite{boj03}. Because of the diagonalization, the basis states will be three copies of those used in the isotropic model. Starting with the eigenstates of the triad 
\[
| m \rangle = \frac{\exp(\frac{imc}{2})}{\sqrt{2} \sin(\frac{c}{2})}
\]
where $m$ is a real number, a gauge-invariant state $| s \rangle$ can be expanded as a series
\[
| s \rangle = \sum_{m_1, m_2, m_3} s_{m_1, m_2, m_3} | m_1 \rangle \otimes | m_2 \rangle \otimes | m_3 \rangle,
\]
where the coefficients $s_{m_1, m_2, m_3}$ satisfy the conditions
\begin{align*}
s_{m_1, m_2, m_3} &= s_{-m_1, -m_2, m_3} = s_{-m_1, m_2, -m_3} \\
&= s_{m_1, -m_2, -m_3}.
\end{align*}
These latter conditions come from the remaining freedom to change the sign of two of the triad vectors simultaneously; it allows us to choose one of the three parameters to run from $-\infty$ to $\infty$, and restrict the other two to non-negative values. For example, if we let $m_3 \in (-\infty, \infty)$, then $m_3 = 0$ would correspond to the classical singularity. In this work, however, we will look only at the subset where all three parameters are non-negative; as we will comment below, the symmetry of the Hamiltonian constraint would allow one to build up a solution for all possible values of $m_k$.

When this classical structure is carried over into quantum operators, we get a partial difference equation for the Hamiltonian constraint in terms of the modified coefficients $t_{m_1, m_2, m_3} = V_{2m_1, 2m_2, 2m_3} s_{2m_1, 2m_2, 2m_3}$, where $V_{m_1, m_2, m_3} = (\frac{1}{2} \gamma \ell _P ^2)^{3/2} \sqrt{| m_1 m_2 m_3 |}$ are the eigenvalues of the volume operator $\hat V$ acting on the state $| m_1 \rangle \otimes | m_2 \rangle \otimes | m_3 \rangle$, and $\ell_P$ is the Planck length. Because of their definition, with $t_{m_1, m_2, m_3}$ including a volume factor, we have that, e.g. $t_{0, m_2, m_3} = 0$, and similarly for the other boundaries. We can use the sequence $t$ since the original values $s_{0, m_2, m_3}, s_{m_1, 0, m_3}$ and $s_{m_1, m_2, 0}$ fall out of the recursion relation and thus have no effect on the final solution. Also, the parameters $m_k$ are all real numbers, but the difference equation relates only those separated by a jump of one step. Here, we will examine those sequences that pass through the classical singularity; the rest of the sequence can be built up by continuity. Because the intermediate values will be interpolated, there are no problems scaling the parameters $m_k$ when mapping between the $s$ and $t$ sequences.
If we define the difference operator $\delta_1$ as
\[
\delta_1 t_{m_1, m_2, m_3} \equiv t_{m_1 + 1, m_2, m_3} - t_{m_1 - 1, m_2, m_3},
\]
and similarly for $\delta_2$ and $\delta_3$, the Hamiltonian constraint in Bianchi I for the particular case of a cosmological constant $\Lambda$ is given by the difference equation
\begin{align}
\nonumber
[ d(m_1) & \delta_2 \delta_3 + d(m_2) \delta_1 \delta_3 + d(m_3) \delta_1 \delta_2 \\
\label{rec-rel}
& + 2 \kappa \gamma^3 \ell^2 _p \Lambda] t_{m_1, m_2, m_3} = 0,
\end{align}
where 
\[
d(n) = \sqrt{ \biggl|1 + \frac{1}{2n} \biggr|} - \sqrt{ \biggl|1 - \frac{1}{2n} \biggr|}
\]
Partial difference equations are more difficult to solve than recursion relations of one parameter, so next we discuss how to find solutions.

In order to find a sequence $t_{m_1, m_2, m_3}$ that satisfies the recursion relation $(\ref{rec-rel})$, we will use the method of generating functions, briefly described here. For more details using this technique in loop quantum cosmology, see~\cite{car-kha-boj04}. We will define a function of three variables such that
\[
F(x, y, z) = \sum_{m_1 = 0} ^\infty \sum_{m_2 = 0} ^\infty \sum_{m_3 = 0} ^\infty t_{m_1, m_2, m_3} x^{m_1} y^{m_2} z^{m_3},
\]
and find a partial differential equation for $F(x, y, z)$ that is equivalent to the recursion relation for $t_{m_1, m_2, m_3}$. It is important to realize that finding a solution $F$ gives the {\it entire} sequence $t$ -- there is no need to choose a particular parameter $m_k$ as a "time" to evolve in, avoiding such issues of interpretation. Notice that we only include positive powers of the variables $x, y$ and $z$ in the series, whereas one of the parameters, say, $m_3$, can assume negative values, passing through the singularity $m_3 = 0$. Because of the symmetry of the recursion relation $(\ref{rec-rel})$ under $m_3 \to - m_3$ from the fact that $d(-m_3) = -d(m_3)$, we can match any two of our solutions at this singularity.

The question now is how to pass between a recursion relation for a sequence, and a condition on the associated generating function. Suppose we have a one-parameter sequence $r_k$ and corresponding generating function $R(x)$. Multiplication by a power $x^d$ will result in shifting the sequence $r_k \to r_{k + d}$, while multiplication by a function of the parameter $k$, e.g. $k r_k$, is related to differentiation by $x$~\cite{wil93}. By using ideas of this type, we can find a differential equation for $F(x, y, z)$ which has the same information as the recursion relation (\ref{rec-rel}). However, the {\it generic} solution found by this method will not be pre-classical, but instead will oscillate in sign as one moving towards increasing $m_k$; we must invoke additional conditions on the function $F$ so that we only get physical solutions.

To do this requires some thought about which sequences will avoid sign oscillation. It turns out that this depends crucially on the pole structure of the function $F$. As an example, a simple pole at $x=1$ will lead to a Taylor series expansion whose coefficients become constant asymptotically, since $(x - 1)^{-1} = 1 + x + x^2 + x^3 + \cdots$. On the other hand, we wish to avoid simple poles at $x=-1$, since the coefficients of the Taylor series alternate sign, with $(x + 1)^{-1} = 1 - x + x^2 - x^3 + \cdots$. So having a pole at $x=1$ will result in a sequence whose coefficients are asymptotically constant. However, to avoid the sequence growing without bound, this must be a simple pole; this behavior can be seen in the expansion of the double pole $(x - 1)^{-2} = 1 + 2x + 3x^2 + 4x^3 + \cdots$. Again, because we are able to look at the {\it global} features of the sequence $t_{m_1, m_2, m_3}$ by fixing the properties of the function $F$, we do not have to postulate a particular wave form that the sequence must "evolve" to at large values of a parameter $m_k$. Instead, we can stipulate that the sequence must asymptotically converge on a solution which has the desired pre-classical properties; the model itself tells us what kinds of asymptotics are possible.

With these ideas, we start with the Hamiltonian constraint $(\ref{rec-rel})$; however, because the function $d(n)$ is not polynomial, we cannot easily change functions of $m_k$ into differential operators. So we make an approximation to allow us to find a differential equation for the generating function; using the fact that $d(n) \simeq 1/2n$, we can write the recursion relation (\ref{rec-rel}) as
\begin{align}
\nonumber
(m_1 m_2 & \delta_1 \delta_2 + m_1 m_3 \delta_1 \delta_3 + m_2 m_3 \delta_2 \delta_3 \\
\label{rec-rel2}
& + 2 \kappa \gamma^3 \ell^2 _p \Lambda) t_{m_1, m_2, m_3} = 0.
\end{align}
Since $d(0) = 0$, this equation works only for $m_1, m_2, m_3 \ge 1$. Because we are making an approximation for the function $d(n)$, we should have terms on the right-hand side of this recursion relation with factors of the difference $[d(m_k) - 1/2m_k]$. However, the size of the error if these terms are neglected will be small compared to the sequence itself~\cite{car-kha-boj04}.

By using the mapping between the recursion relation and partial differential equations discussed above, we can find an equation for a three-variable function $F(x, y, z)$, whose solutions correspond to sequences $t_{m_1, m_2, m_3}$ that solve the approximate recursion relation $(\ref{rec-rel2})$ for the Bianchi I model. This is given by
\begin{align}
\nonumber
& \frac{\partial^2}{\partial x \partial y} \biggl[ \frac{(1 - x^2)(1 - y^2) F}{xyz} \biggr]
+ \frac{\partial^2}{\partial x \partial z} \biggl[ \frac{(1 - x^2)(1 - z^2) F}{xyz} \biggr] \\
\label{gen-eqn}
& + \frac{\partial^2}{\partial y \partial z} \biggl[ \frac{(1 - y^2)(1 - z^2) F}{xyz} \biggr] =
- 2 \kappa \gamma^3 \ell^2 _p \Lambda \frac{\partial^3 (xyz F )}{\partial x \partial y \partial z}.
\end{align}
With this equation, it is natural to define a new function $H(x, y, z)$, such that
\[
F(x, y, z) = \frac{x y z H(x, y, z)}{(1 - x^2)(1 - y^2)(1 - z^2)}.
\]
Not only does this make the differential equation $(\ref{gen-eqn})$ for the generating function more tractable, but it also gives us a way to ensure that the function has the proper asymptotic behavior. By assuming that the function $H(x, y, z)$ has boundary conditions $H(-1, y, z) = H(x, -1, z) = H(x, y, -1) = 0$, the original function $F(x, y, z)$ can have at most a simple pole at $x = -1$, etc.  {\it The requirement for a pre-classical sequence $t_{m_1, m_2, m_3}$ is manifest as a set of boundary conditions for the derived function $H$}. When we substitute for $F$ in terms of $H$, and multiply by $(1 - x^2)(1 - y^2)(1 - z^2)$, this gives us the equation
\begin{widetext}
\begin{align}
& (1 - x^2)(1 - y^2) \frac{\partial^2 H}{\partial x \partial y} + (1 - x^2)(1 - z^2) \frac{\partial^2 H}{\partial x \partial z} + (1 - y^2)(1 - z^2) \frac{\partial^2 H}{\partial y \partial z} \\
\nonumber
& = - 2 \kappa \gamma^3 \ell^2 _p \Lambda (1 - x^2)(1 - y^2)(1 - z^2) \frac{\partial^3}{\partial x \partial y \partial z} \biggl[ \frac{x^2 y^2 z^2 H}{(1 - x^2)(1 - y^2)(1 - z^2)} \biggr],
\end{align}
\end{widetext}
with boundary conditions mentioned above. We show next that, regardless of the value of the cosmological constant $\Lambda$, this equation only has the trivial solution $H(x, y, z) = 0$.

Because of the boundary conditions that are necessary to assure pre-classicality, any solution $H(x, y, z)$ will be the sum of terms in the form $(x + 1)^i (y + 1)^j (z + 1)^k$. We will use this fact to show that the case $\Lambda = 0$ has only trivial pre-classical solutions. First, we define new variables $u = x + 1, v = y + 1,$ and $w = z + 1$, and look at the lowest order terms in $u, v$ and $w$, namely
\[
H(u, v, w) = \sum_{i + j + k = d} c_{i, j, k} u^i v^j w^k + \cdots,
\]
with the parameter $d$ defining the order of the term. Note that, due to the boundary conditions, each of the individual powers $i, j, k \ge 1$. To lowest order in these new variables, the partial differential equation for $H(u, v, w)$ is
\[
uv \frac{\partial^2 H}{\partial u \partial v} + uw \frac{\partial^2 H}{\partial u \partial w} + vw \frac{\partial^2 H}{\partial v \partial w} = 0
\]
so that, when we put our lowest order terms in the equation, the result is
\[
 \sum_{i + j + k = d} (ij + ik + jk) c_{i, j, k} u^i v^j w^k = 0.
 \]
Since none of the exponents $i, j$ or $k$ are negative to ensure that the solution has no additional poles -- the very definition of $H(u, v, w)$ ensures we have just enough to give a pre-classical solution -- then we conclude that the coefficients $c_{i, j, k} = 0$. Since this argument works regardless of the value of the lowest order parameter $d$, the only solution we have to the differential equation with these boundary conditions is $H(u, v, w) = 0$. Therefore, {\it the only pre-classical solution to the $\Lambda = 0$ Bianchi I model is $t_{m_1, m_2, m_3} = 0$}. When we include a non-zero cosmological constant, the equation to solve becomes more complicated, and there is no simple way to show that there is only the null solution. In this case, there is an inconsistency between the coefficients of different orders in the series expansion of $H(u, v, w)$, and so the only solution for the $\Lambda \ne 0$ situation is also $c_{i, j, k} = 0$.

These results are contrary to the expectation that the less constrained Bianchi I model would have at least as many, if not more, solutions than the isotropic case. One reason this does not occur may be in the chosen approach to quantization; a different method for quantizing the Hamiltonian constraint may allow more pre-classical solutions. For example, one can consider an alternate factor ordering when the constraint is quantized.  In all the work done previously, such as the isotropic models, the triad  operators are placed on the right~\cite{boj02, thi98}. The problems seen in this present work may not carry over to another ordering, say a symmetric one. However, since we are considering a reduction of the full loop quantum gravity situation, one should use one choice of ordering consistently for all such symmetry reductions. If it turns out that our current choice is not the best for a specific model, this must be considered in the results of all other cases. This places a meaningful restriction on what methods can be used in the full theory of quantum geometry.

On the other hand, it is possible that there are no technical issues of this sort, and loop quantum cosmology is only feasible on certain types of space-times. Here, in the full Bianchi I model, there are no pre-classical solutions whatsoever. When we examine the Bianchi I LRS case~\cite{car-kha-boj04}, pre-classical states have been found, but they are rather limited. Near the singularity, the solution will assume a wave form that remains constant as the two parameters $m, n$ increase in value (here, $m$ labels the two equal triad components, $n$ the third). In fact, the only variation that occurs in the solution is in the $n$ direction; the sequence quickly assumes a constant value as one moves in the $m$ direction of parameter space. Thus, for the two anisotropic cases considered so far where full quantum solutions are known, there are strong restrictions on the possible states. It is possible that the inclusion of matter in the Bianchi I model may change this situation; a cautionary note would be that this does not occur in the closed isotropic model~\cite{gre-unr04}. It remains to be seen if these types of restrictions are solely for Bianchi I, or whether they occur more generically in anisotropic space-times. 

The authors appreciate the helpful suggestions of Martin Bojowald, Robert Israel and Jorge Pullin in writing this manuscript. GK is grateful for research support from the University of Massachusetts at Dartmouth, as well as Glaser Trust.


\begin{thebibliography}{99}

\bibitem{lqc} A. Ashtekar, M. Bojowald and J. Lewandowski, Adv. Theor. Math. Phys. {\bf 7}, 233 (2003); M. Bojowald, Pramana {\bf 63}, 765 (2004)

\bibitem{rov} C. Rovelli, {\it Quantum Gravity} (Cambridge University Press, Cambridge, 2004)

\bibitem{boj01} M. Bojowald, Phys. Rev. Lett. {\bf 87}, 121301 (2001); M. Bojowald, Gen. Rel. Grav. {\bf 35}, 1877 (2003)

\bibitem{boj02} M. Bojowald, Class. Quantum Grav. {\bf 19}, 2717 (2002)

\bibitem{inf} M. Bojowald, Phys. Rev. Lett. {\bf 89}, 261301 (2002)

\bibitem{osc} P. Singh and A. Toporensky, Phys. Rev. D {\bf 69}, 104008 (2004); J. E. Lidsey, D. J. Mulryne, N. J. Nunes and R. Tavakol, Phys. Rev. D {\bf 70}, 063521 (2004); G. Date and G. M. Hossain, Phys. Rev. Lett. (to appear), gr-qc/0407074

\bibitem{semi} G. Date and G. M. Hossain, Class. Quantum Grav. {\bf 21}, 4941 (2004)

\bibitem{boj-dat-hos04} M. Bojowald, G. Date and G. M. Hossain, Class. Quantum Grav. {\bf 21}, 3541 (2004)

\bibitem{car-kha-boj04} D. Cartin, G. Khanna and M. Bojowald, Class. Quantum Grav. {\bf 21}, 4495 (2004)

\bibitem{wil93} H. Wilf, {\it Generatingfunctionology} (Academic, New York, 1993)

\bibitem{mal04} D. Malecki, Phys. Rev. D {\bf 70}, 084040 (2004)

\bibitem{boj03} M. Bojowald, Class. Quantum Grav. {\bf 20}, 2595 (2003)

\bibitem{thi98} T. Thiemann, Class. Quantum Grav. {\bf 15}, 839 (1998)

\bibitem{gre-unr04} D. Green and W. G. Unruh, Phys. Rev. D {\bf 70}, 103502 (2004)

\end{thebibliography}
\end{document}